\documentclass[aps,preprint,showpacs]{revtex4}
\usepackage{graphicx}

\begin{document}
\title{A Correction Method for the Density of States}
\author{Shijun Lei}
\affiliation{Purple Mountain Observatory, 2 West Beijing Rd, Nanjing Jiangsu, 210008 P.R. China}
\date{\today}

\begin{abstract}
We present a correction method for the density of states (DOS) obtained from the generalized ensemble simulations. The DOS
is proportionally corrected to match the exact values and/or good approximations known for the system. We demonstrate the
validity of the method by applying it to the DOS of 2D Potts model calculated from various generalized ensemble simulations.
It is shown that the root-mean-square error of the DOS is reduced by $\sim50\%$ or more without additional heavy calculations.

\end{abstract}
\maketitle
Generalized ensemble Monte Carlo (MC) algorithms, such as the multicanonical method \cite{Berg91}, the Wang-Landau
method \cite{Wang02}, and variations thereof \cite{Lee93,Yan03,Shell03,Troyer03,Belardinelli07}, are effective simulational
tools for probing complex systems that are often characterized by their rough energy landscape. Unlike the traditional Metropolis
method that simulates canonical ensembles at distinct temperatures, these methods simulate a flat histogram ensemble
to calculate the density of states (DOS) of a wide energy range. Previously, efforts have been mainly focused on optimizing the
performance of the simulation: either to change the dynamics \cite{Yamaguchi02,Berg06} or to change the simulated statistical
ensemble \cite{Trebst04} of the simulation. We show here a different approach to achieve higher accuracy and efficiency.
The DOS could be proportionally corrected to match the exact values and/or good approximations known for the system
to improve the overall accuracy without heavy calculations. The validity of the method is demonstrated by applying it to the DOS
of 2D Potts model calculated from multicanonical and $1/t$ WL \cite{Belardinelli07} simulations. 

The errors in the absolute DOS for a Q=8 2D Potts model of lattice size $N=8\times8$ (hereafter ``the Potts model") calculated
from a $1/t$ WL simulation of $t=10^7$ MC moves (flips) is shown in Fig. 1 as the solid curve. Since the simulation provides only
a relative DOS for different energies, the absolute DOS is extracted using the fact that the number of ground stats (where $E_1=-2N$)
is $Q$ and is hereafter referred to as DOS for simplicity. Throughout this paper, we use the absolute error defined as
$\varepsilon(E)=S(E)-S_e(E)=\ln[g(E)]-\ln[g_e(E)]$ in comparing the simulational ($g(E)$) and the exact ($g_e(E)$)
DOS values. Here the ``exact" DOS ($S_e(E)=\ln[g_e(E)]$) is calculated from a long run of $1/t$ Wang-Landau simulation of
$t=10^{12}$ flips and we have tested that our discussion below is not affected by the errors remained in $S_e(E)$. One may note
that the curve differs from a white noise that has no correlation between different values. Actually, a correlation between the errors
is explicitly revealed in Fig. 2 where the correlation coefficient calculated as
$R_{\varepsilon}(\Delta E)=\sum_E\varepsilon(E)\varepsilon(E+\Delta E)$ is plotted as a function of the energy difference
($\Delta E$). For calculating the correlation coefficient, the DOS is normalized in a way that the average of the errors is zero. 
This correlation between the errors suggests a possibility to correct the whole DOS using the knowledge about
certain errors.

For many systems, more than one exact DOS values and/ or good approximations are known. Actually, besides
the rescaling condition used above, the DOS of the Potts model could also be normalized using the fact that the total number
of possible states is $\sum_Eg_n(E)=Q^N$ \cite{Wang02}. While the first rescaling condition ($\ln[g_n(E)]=\ln[g(E)]-\ln[g(E_1)]+\ln(Q)$)
guarantees the accuracy of the DOS at low energy levels, the second one ($\ln[g_n(E)]=\ln[g(E)]-\ln[\sum_Eg(E)]+N\ln[Q]$) ensures the
accuracy of the maximum DOS values as the summation $\sum_Eg(E)$ is dominated by those values. (We denote the maximum DOS
value as $g(E_n)$.) The discrepancy between these two rescaling conditions,
$\epsilon(E_n)-\epsilon(E_1)=(N\ln[Q]-\ln[\sum_Eg(E)])-(\ln(Q)-\ln[g(E_1)])$, gives the absolute error in the maximum DOS values.
With the knowledge of the error $\epsilon(E_1)$ and $\epsilon(E_n)$, we propose a linear correction to all the DOS values $S(E_i)$
( $i=1,2,3...n$) within the energy range $[E_1,E_n]$,
\begin{equation}
S_c(E_i)=S(E_i)+\varepsilon(E_1)+[\varepsilon(E_n)-\varepsilon(E_1)]/(n-1)\times(i-1)
\end{equation}
We are goingto show that this simple correction that stretches or compresses the DOS curve proportionally to match the exact values
improves the overall accuracy of the DOS.

Shown in Fig. 3 are the root-mean-square (rms) error in both the corrected and non-corrected DOS of the Potts model as a function of
simulation time. Here the rms error calculated as $\sigma(E)=\sqrt{\frac{\sum_i[S(E_i)-S_e(E_i)]^2}{n}}$ is a benchmark of the errors
in the $n$ DOS values within the energy range of $[E_1,E_n]$. To put the comparison on the same base, the non-corrected DOS is
normalized using the same $\varepsilon(E_1)$. We find that in average the correction reduces the rms error by $50\%$, independent
to the simulation time and model size. In other words, this correction could save $75\%$ of the simulation time to achieve the same
overall accuracy, given that $\sigma\propto1/\sqrt{t}$. Obviously, the correction could be extended straightforwardly to the case that
even more exact DOS values are known. As a simple test, for the DOS of a 2D Ising model of various size, we achieve a reduction in
rms error by $65\%$ by applying the correction to both halves of the DOS around the peak.

To reduce the error caused by the local fluctuations in the simulated DOS and make $\varepsilon(E)$ more representative of the overall
error that we should correct, the average of the errors in several neighboring $g(E)$ could be used. As in the case above, the error
calculated from $\varepsilon(E_n)=N\ln[Q]-\ln[\sum_Eg(E)]$ automatically makes use of the several maximum DOS values that are
significant to determine the total number of possible states and is thus a better choice than that simply calculated as
$\varepsilon(E_n)=g_e(E_n)-g(E_n)$. We remark that $g_e(E_n)$ could be calculated to sufficient accuracy using a Metropolis simulation
of the system at infinite temperature with slight calculations, and could be used together with $g_e(E_1)$ to make a better initial guess for
the simulations.) Also, it is trivial to find out for the Potts model the $g(E)$ of the first, second, and third excited energy levels,
$g(E_2=-2N+4)=NQ(Q-1)$, $g(E_3=-2N+6)=2NQ(Q-1)$, and $g(E_4=-2N+7)=2NQ(Q-1)(Q-2)$. we thus use
$\overline{\varepsilon}(E_1)=[\varepsilon(E_1)+\varepsilon(E_2)+\varepsilon(E_3)+\varepsilon(E_4)]/4$ in our correction.

A more direct test of the validity of the correction might be comparing the accuracy of the thermaldynamic values calculated from the
corrected and non-corrected DOS. For the Potts model, we show in Fig. 4 the histogram of the temperatures of the peak of the specific
heat ($T_c$) obtained from 100,000 realizations of malticanonical and $1/t$ WL simulation. For both simulations the correction is more
effective than doubling the simulation time. The gain in the efficiency for calculating $T_c$, however, is not so large as that we find for
$\sigma(E)$. This is because the correction that improves the overall accuracy of the DOS is more effective to the value that depends
on ``long-range" accuracy of the DOS. And note that while the $\sigma(E)$ is calculated for the entire corrected DOS curve, the critical
temperature is only determined by about half of the curve \cite{Landau02}. As noted before \cite{Caparica12}, the DOS resulted from
$1/t$ WL simulation has systematic errors that cause the deviation of the center of the distribution from the exact value. For this case,
not only the correction sharpen the distribution but also shift the center toward the correct value significantly, showing that the correction
is also effective to reduce the systematic errors in the DOS. We want to point out that the better results from the multicanonical simulations
is because it uses a constant weighting $g_i(E)$ obtained from a $1/t$ WL simulation using $2\time10^6$ flips, while the $1/t$ WL
simulations start from a initial guess $g(E)=0$. 

In summary, we propose a correction method for the DOS calculated from generalized ensemble simulation. The DOS could be
corrected to match the exact values and/or good approximations known for the system to improve the overall accuracy. Applying
the method to 2D Potts model and Ising model, we achieve a reduction in the rms error by $50\%$ and $65\%$, respectively.
Comparing to the simulation effort, this correction hardly costs any time.

\begin{figure}
\centering{\includegraphics[angle=90,width=1.1\textwidth]{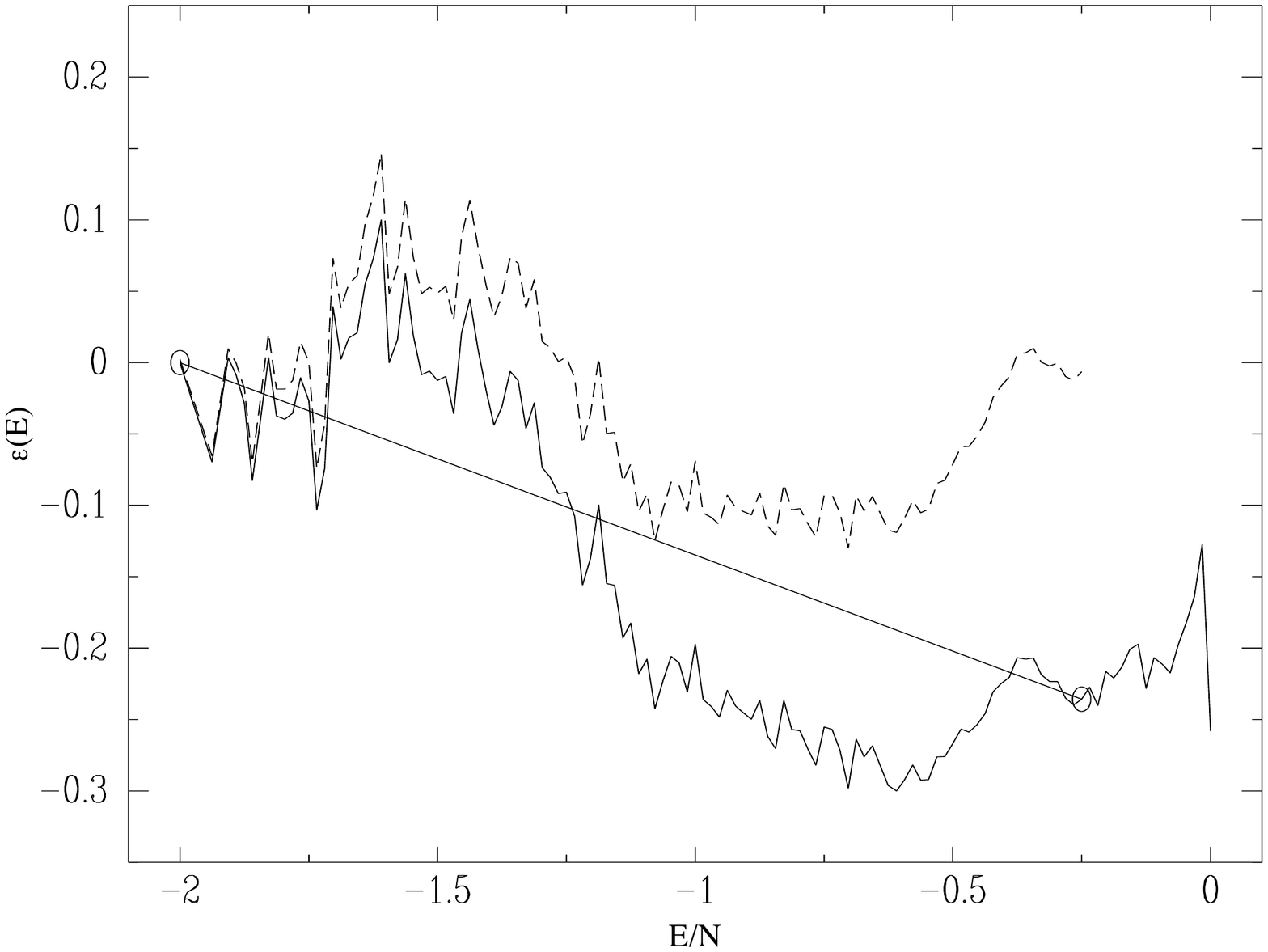}}
\centering\caption{Errors in a simulated DOS of the $8\times8$ Q=8 Potts model as a function of energy per lattice site, $E/N$
(solid curve). The DOS is calculated by a $1/t$ WL simulation using a single run of $10^7$ MC moves and is normalized using
the fact that the number of ground states is $Q$. The errors after correcting the straight line connecting the two error points
($E/N=-2, -0.25$ and marked as circles) whose values are known is shown as the dashed curve within the energy range [-2,-0.25].}
\label{fig1}
\end{figure}

\begin{figure}
\centering{\includegraphics[angle=270,width=1.1\textwidth]{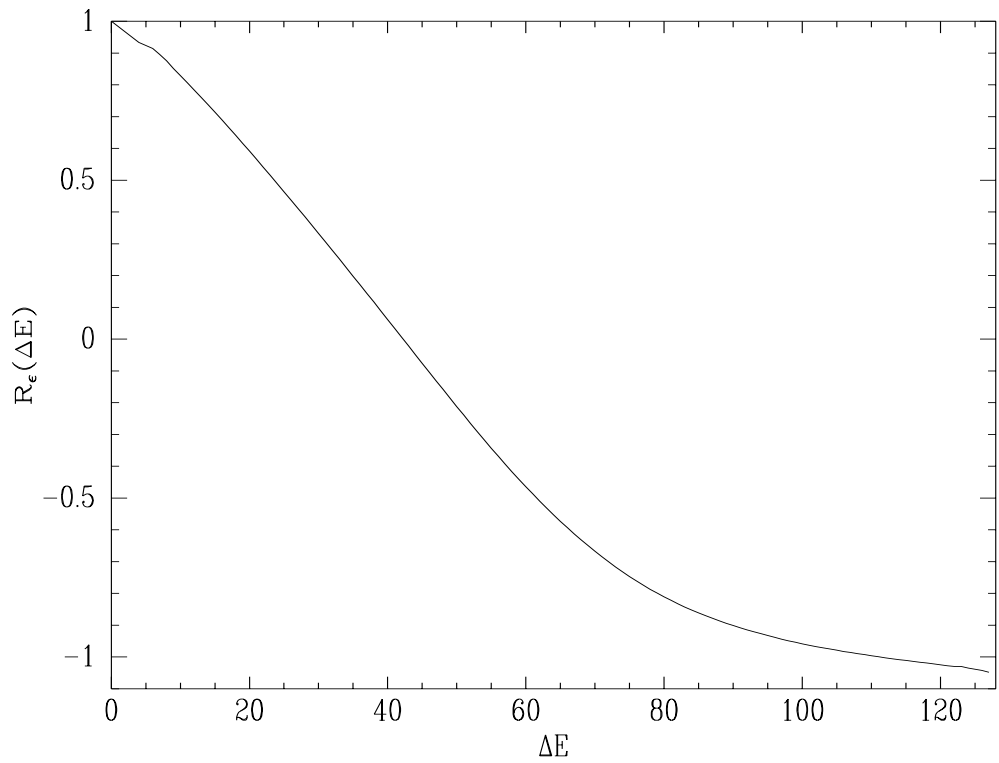}}
\centering\caption{Correlation coefficient of the errors in the DOS of the Potts model as a function of energy difference.
The coefficient is calculated using 10,000 DOS obtained from independent $1/t$ WL simulations of $t=10^7$ flips}
\label{fig2}
\end{figure}

\begin{figure}
\centering{\includegraphics[angle=0,width=1.1\textwidth]{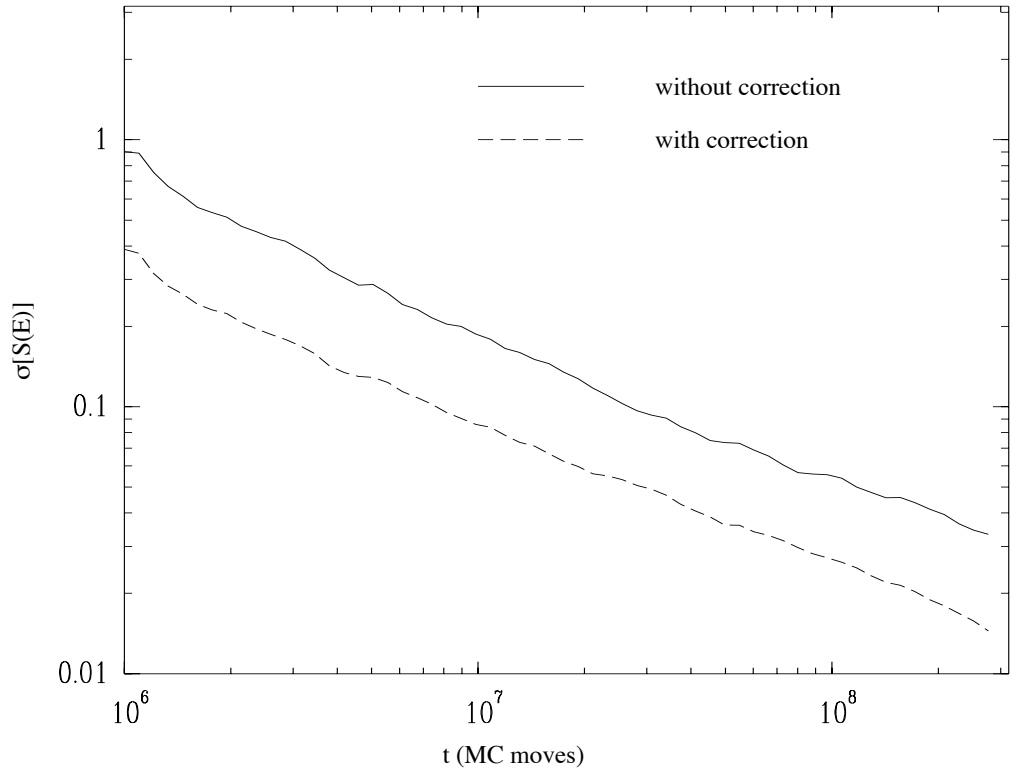}}
\centering\caption{The rms error of the DOS of the Potts model as a function of multicanonical simulation time. Each curve is the
average of 100 realizations.}
\label{fig3}
\end{figure}

\begin{figure}
\centering{\includegraphics[angle=270,width=1.1\textwidth]{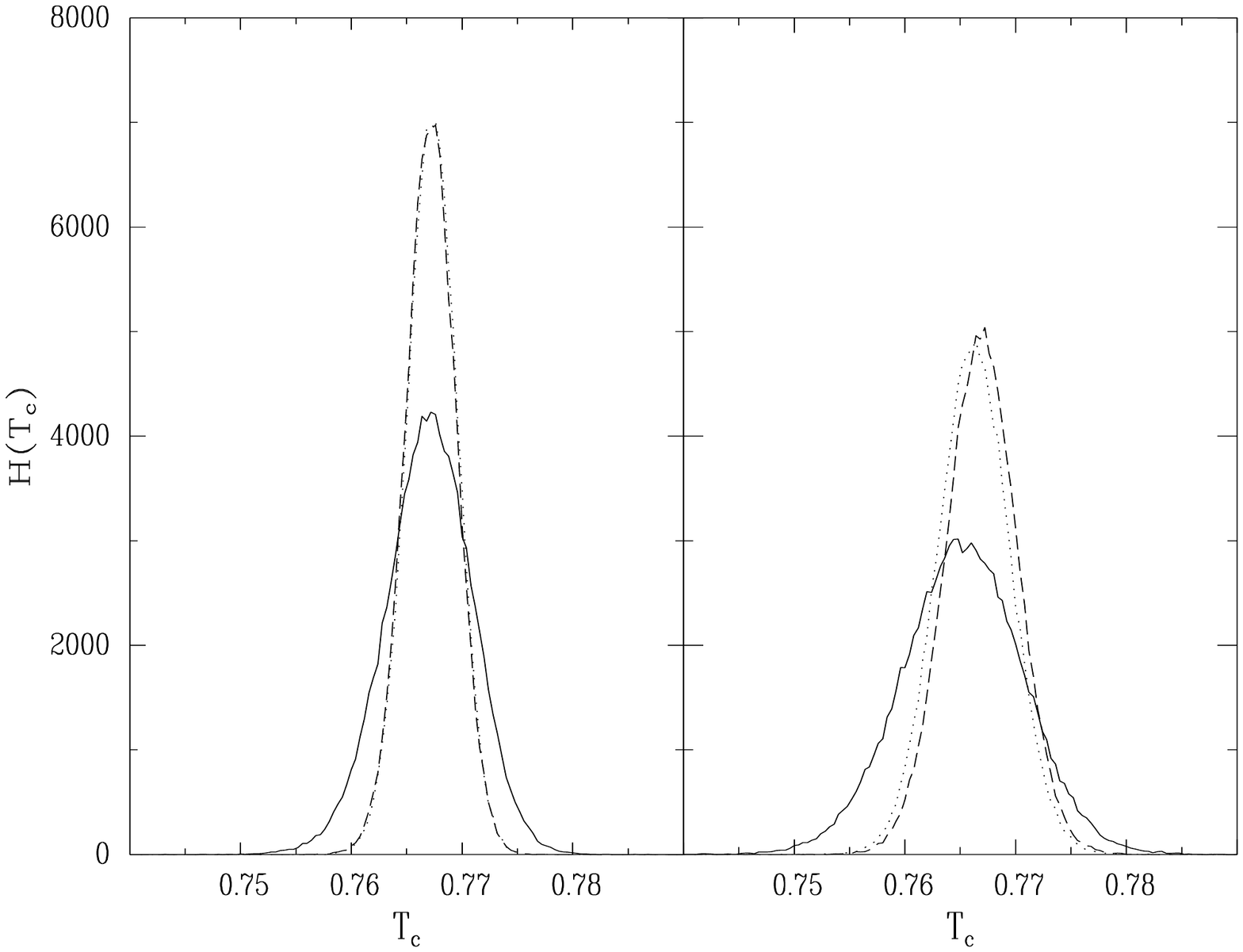}}
\centering\caption{Left panel: Histograms of the location of the peak of the specific heat for the Potts model resulted from 100,000
realizations of multicanonical simulations. The solid curve is calculated from the non-corrected DOS obtained by simulations using
$2\times10^6$ flips. The dashed and dotted curve (almost overlapping with each other) correspond to applying the correction and
tripling the simulation time to $6\times10^6$ flips, respectively. Right panel: Same as the left panel but for $1/t$ WL simulations. The
dotted curve correspond to doubling the simulation time to $4\times10^6$ flips. See text for more details.}
\label{fig4}
\end{figure}


\begin{thebibliography}{99}

\bibitem{Berg91} B. A. Berg and T. Neuhaus, Phys. Lett. B {\bf 267}, 249 (1991);
Phys. Rev. Lett. {\bf 68}, 9 (1992).

\bibitem{Wang02} F. Wang and D. P. Landau, Phys. Rev. Lett. {\bf 86}, 2050 (2001); Phys. Rev. E {\bf 64}, 056101 (2001); 

\bibitem{Lee93} J. Lee, Phys. Rev. Lett., {\bf 71}, 211 (1993).

\bibitem{Yan03} Q. Yan and J. J. de Pablo, Phys. Rev. Lett. {\bf 90}, 035701 (2003).

\bibitem{Shell03} M. S. Shell et al., J. Chem. Phys. Rev. {\bf 119}, 9406 (2003).

\bibitem{Troyer03} M. Troyer et al., Phys. Rev. Lett. {\bf 90}, 120201 (2003).

\bibitem{Belardinelli07} R. E. Balardinelli, and V. D. Pereyra, J. Chem. Phys. {\bf 127}, 18 (2007).

\bibitem{Yamaguchi02} C. Yamaguchi and N. Kawashima, Phys. Rev. E {\bf 65}, 056701 (2002).

\bibitem{Berg06} B. A. Berg and W. Janke, Phys. Rev. Lett. {\bf 90}, 040602 (2007).

\bibitem{Dayal04} P. Dayal, S. Trebst, S. Wessel, D. Wurtz, M. Troyer, S. Sabhapandit and S. N. Coppersmith,
Phys. Rev. Lett. {\bf 92}, 097201 (2004).

\bibitem{Trebst04} S. Trebst and D. A. Huse and M. Troyer, Phys. Rev. E {\bf 70}, 046701 (2004).

\bibitem{Landau02} D. P. Landau and Shan-Ho Tsai and M. Exler, Comput. Phys. Commun. {\bf 147}, 674 (2002).

\bibitem{Caparica12} A. A. Caparica and A. G. Cunha cond-mat,stst-mech/.1110.4517v2

\end{thebibliography}
\end{document}